\documentclass[twocolumn]{aastex631}

\usepackage{amsmath, amssymb}
\usepackage{graphicx}
\usepackage{booktabs}
\newcommand{\solarmass}{\>{\rm M_{\odot}}}
\usepackage{hyperref}

\begin{document}

\title{Evolution of Cluster Alignments as Evidence of Large-scale Structure Formation in the Universe}

\correspondingauthor{Michael J. West}
\email{mwest@lowell.edu}

\correspondingauthor{J.L. Han}
\email{hjl@bao.ac.cn}

\author[0000-0002-6091-6083]{Michael J. West}
\affiliation{Lowell Observatory,
1400 W. Mars Hill Rd.,
Flagstaff, AZ, 86001 USA}

\author[0000-0003-1455-7339]{Roberto De Propris}
\affiliation{Finnish Centre for Astronomy with ESO,
University of Turku,
Vesilinnantie 5, Turku FI-21400, Finland}
\affiliation{Department of Physics and Astronomy,
Botswana International University of Science \& Technology, Private Bag 16,
Palapye, Botswana}

\author[0000-0003-3722-8239]{Maret Einasto}
\affiliation{Tartu Observatory, 
University of Tartu, Observatooriumi 1, 61602 
T\~oravere, Estonia}

\author[0000-0003-3722-8239]{Z.L. Wen}
\affiliation{National Astronomical Observatories, Chinese Academy of Sciences, A20 Datun Road, Chaoyang District, Beijing 100101, China}
\affiliation{CAS Key Laboratory of FAST, NAOC, Chinese Academy of Sciences}

\author[0000-0002-9274-3092]{J.L. Han}
\affiliation{National Astronomical Observatories, Chinese Academy of Sciences, A20 Datun Road, Chaoyang District, Beijing 100101, China}
\affiliation{CAS Key Laboratory of FAST, NAOC, Chinese Academy of Sciences}
\affiliation{School of Astronomy and Space Sciences, University of Chinese Academy of Sciences, Beijing 100049, China}

\begin{abstract}

The universe’s large-scale structure forms a vast, interconnected network of filaments, sheets, and voids known as the cosmic web. For decades, astronomers have observed that the orientations of neighboring galaxy clusters within these elongated structures are often aligned over separations of tens of Mpc. Using the largest available catalog of galaxy clusters, we show for the first time that clusters orientations are correlated over even larger scales $-$ up to 200 - 300 comoving Mpc $-$ 
and such alignments are seen to redshifts of at least $z \simeq 1$. 
Comparison with numerical simulations suggests that 
coherent structures on similar scales may be expected in $\Lambda$CDM models. 
\end{abstract}


\section{Introduction} \label{sec:intro}

Studies of the large-scale distribution of matter in the universe provide insights into the formation and evolution of structure as well as constraints on cosmological parameters such as the primordial spectrum of density fluctuations and the dark energy equation of state
\citep{Bond1996,Tegmark2004,Eisenstein2005,Springel2005,Blake2011}. 

Decades of two- and three-dimensional galaxy surveys have mapped the structure of the low-redshift universe in increasingly fine detail  \citep{Einasto1980,Zeldovich1982,deLapparent1986,Anderson2014,Abbott2018,Einasto2025}, revealing long, luminous strands of galaxies woven together into a vast cosmic web. Nowhere is this more evident than the nearby Perseus-Pisces Supercluster (Fig. 1), a chain of galaxies that snakes across a large swath of the northern sky, fed by a network of smaller filaments that resemble tributaries flowing into a river. Embedded within these filaments are densely populated groups and clusters of galaxies. Between them lie enormous under-dense regions.
Tracing such structures at earlier epochs is challenging, however, because galaxies are fainter at high redshifts and gravity has had less time to amplify the initial density perturbations. 

\begin{figure*}[ht!]
\centering
\epsscale{1.1}
\plotone{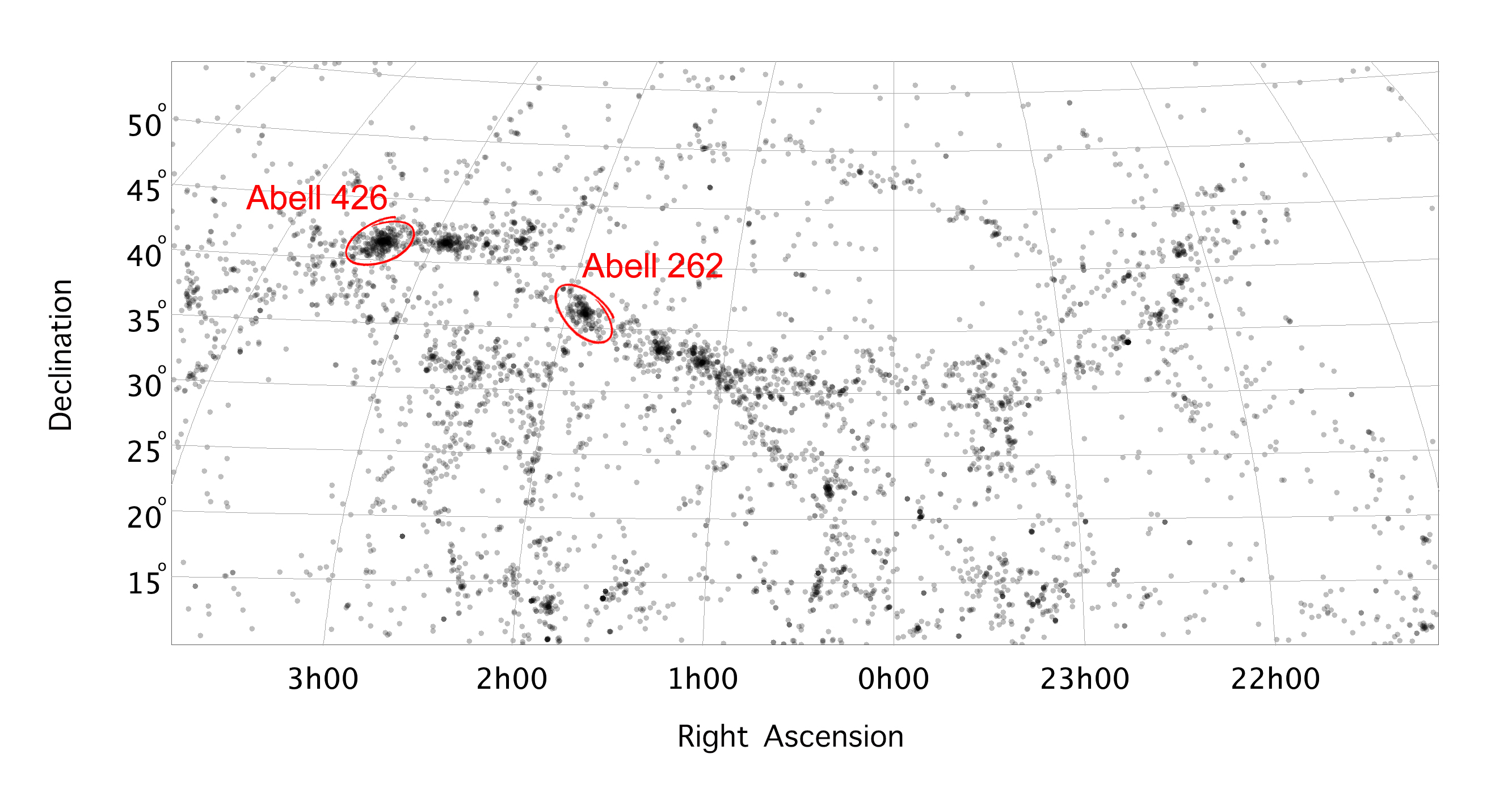}
\caption{The Perseus-Pisces supercluster region is a nearby segment of the cosmic web. Each point in this figure represents a galaxy 
in NASA's Extragalactic Database with a 
recessional velocity in the range $4000-9000$ km/s, corresponding to distances of $\sim 150$ to 300 million light years from Earth.
Densely populated groups and clusters of galaxies dot the prominent filament like beads on a string.
The major axis orientations of two of these clusters, Abell 426 and Abell 262, are shown.} 
\label{fig:PP}
\end{figure*}

As the most luminous gravitationally bound objects in the universe today, 
clusters of galaxies
are sparse but faithful tracers of the cosmic web $-$ not only in their spatial distribution, but also in their orientations. 
Almost half a century ago, \cite{Einasto1980} pointed out that the clusters in the Perseus-Pisces supercluster tend to elongate in the same direction as the filament that bridges them.
These authors suggested that ``the orientations of clusters in superclusters is a conspicuous morphological property of superclusters," and as such could serve as useful probes of the coherence of large-scale structures. Numerous studies have since confirmed this tendency for cluster orientations to be correlated over separations up to $\sim 100$ Mpc\footnote{We assume a standard $\Lambda$CDM cosmology with dark energy ($\Lambda$) and matter (M) density parameters of $\Omega_{\Lambda} = 0.69$ and 
$\Omega_{\rm M} = 0.31$, and a Hubble constant of 67 km/s/Mpc throughout this paper.} or more 
\citep{Binggeli1982,Plionis1994,Struble1985,West1989,Smargon2012,vanUitert2017}. Similar alignments are also seen for smaller groups of galaxies that infall into clusters along supercluster axes \citep{Wang2009,Paz2011,Einasto2018a,Lamman2024}.

However, because clusters are rarer and harder to detect at earlier epochs, studies of cluster alignments have generally been limited to low redshifts, leaving much of the cosmic web's evolution still unexplored. Until now, only \cite{Smargon2012} measured alignments of galaxy cluster pairs to
$z < 0.44$ using two galaxy cluster catalogs extracted from SDSS data. They find an alignment signal to 100 Mpc/$h$ 
but only weak alignments below 20 Mpc/$h$.  With this motivation, we extended the analysis of
cluster alignments to $z \simeq 1.5$ using the recent catalog of \citet{Wen2024}, as described below.

\section{Observations}

\subsection{Cluster sample}
\citet{Wen2024}, hereafter WH24, catalogued 1.58 million galaxy clusters to redshifts $z \simeq 1.5$ based on photometry of 295.6 million galaxies in the 
Dark Energy Spectroscopic Instrument (DESI) Legacy Survey{\footnote{ https://www.legacysurvey.org/}} DR10 and 361.1 million galaxies from DR9 \citep{Dey2019}.
Clusters were identified as overdensities in the distribution of galaxies within redshift slices centered on preselected massive galaxies whose luminosity, stellar mass, and red colors were consistent with being candidate brightest cluster galaxies.
Clusters were then confirmed if the surrounding galaxy population exceeded specific thresholds in richness and member count within $r_{500}$, the radius within which the average density of the cluster is 500 times the critical density of the universe at that redshift.
This is by far the largest sample of galaxy clusters available presently, covering more than 20,000 deg$^2$ of the sky. It
includes most rich optically-selected clusters in previous catalogs, more than 95\% of massive Sunyaev–Zeldovich
clusters, and 90\% of the ROSAT and eROSITA X-ray clusters.
A sky plot of all 1.58 million clusters is shown in Fig. 2, along with the distributions of their redshifts and the number of member galaxies. Cluster equatorial coordinates and redshifts in the WH24 catalog were transformed to a three-dimensional comoving Cartesian coordinate system. 

\begin{figure*}[ht!]
\centering
\epsscale{1.0}
\plotone{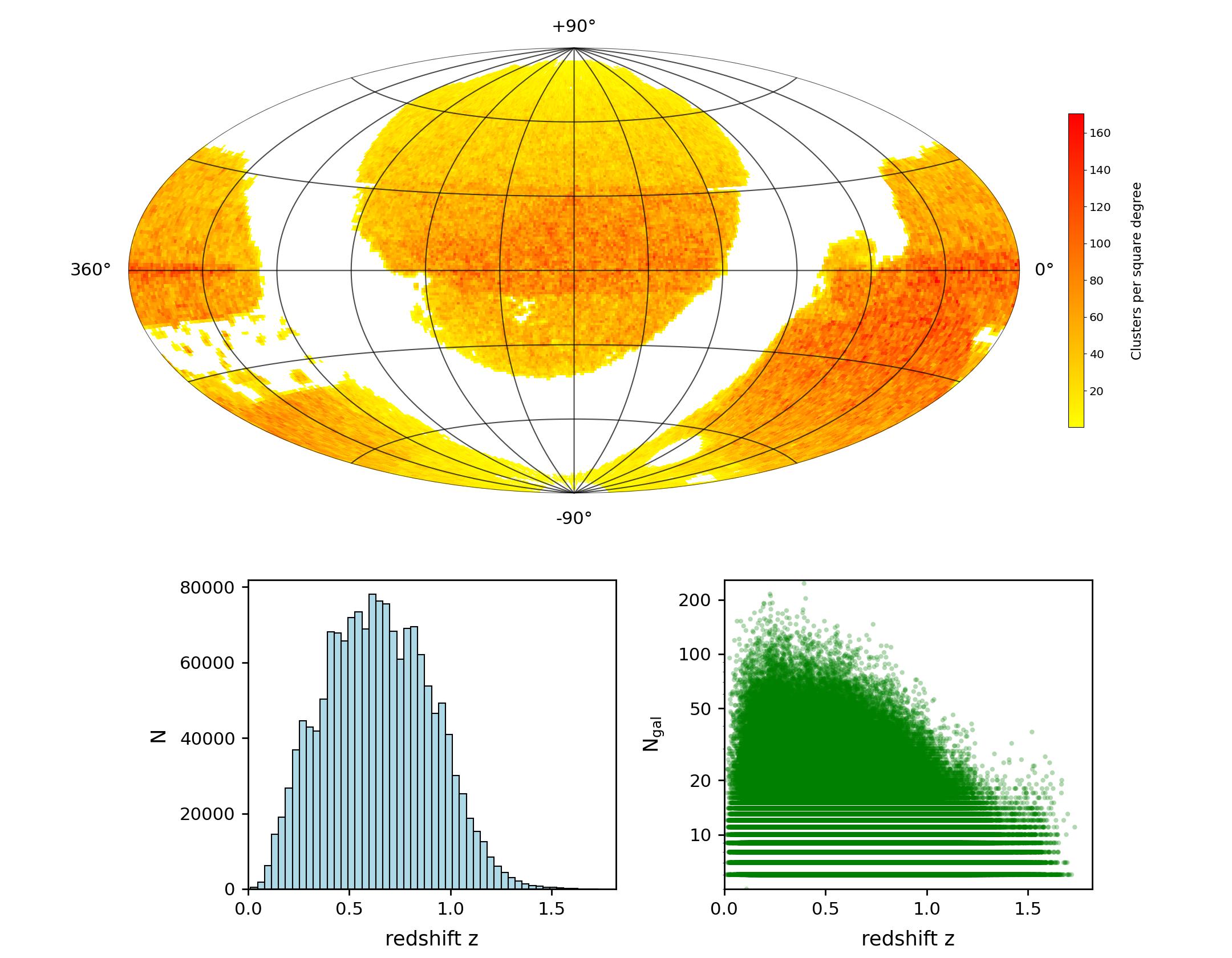}
\caption{Top: A number density plot of the sky distribution of 1.58 million clusters in the Wen \& Han (2024) catalog. Bottom left: The redshift distribution of the clusters. Bottom right:  The number of detected member galaxies in each cluster as a function of redshift. For reference, one degree on the sky subtends roughly 60 cMpc at redshift $z = 1$.}
\label{fig:aitoff}
\end{figure*}

\subsection{Cluster orientations and uncertainties}

The orientation of each cluster, $\phi$, and the ellipticity, $e$, were determined from the projected distribution of its member galaxies in the plane of the sky. This was done by computing moments of inertia of the galaxy distribution with respect to the cluster centroid, which, following WH24, is taken to be the location of the brightest member galaxy. The reduced moment of inertia tensor for a 2D distribution of galaxies is given by:

\[
\tilde{I}_{ij} = \frac{\sum\limits_{k=1}^{N_{\text{gal}}} w_k\, r_{ki} r_{kj}}{\sum\limits_{k=1}^{N_{\text{gal}}} w_k\, r_k^2}
\]

\noindent where:
\begin{itemize}
    \item $w_k$ is the weight of the $k$th galaxy (e.g., unity or luminosity).
    \item $r_{ki}$ and $r_{kj}$ are the $i$th and $j$th components of the \textit{centroid-subtracted} position vector of the $k$th galaxy.
    \item $r_k^2 = r_{k1}^2 + r_{k2}^2$ is the squared distance of the $k$th galaxy from the centroid.
    \item $N_{\text{gal}}$ is the total number of detected galaxies in the cluster.
\end{itemize}

\noindent This tensor is symmetric, dimensionless, and captures the shape and orientation of the projected galaxy distribution independent of physical scale.

The distribution of member galaxies is shown for a small portion of the catalog in Fig. 3.
The WH24 clusters have from six to 246 galaxies, with a median of eight members. The most distant clusters are generally the least populous because only their brightest members are detected, and clusters were less massive at earlier epochs. Although the WH24 catalog becomes increasingly sparse at high redshifts, such incompleteness does not bias our analysis of cluster alignments, except in so far as it makes them harder to detect at earlier epochs.

\begin{figure}[ht!]
\centering
\epsscale{0.8}
\includegraphics[width=0.45\textwidth]{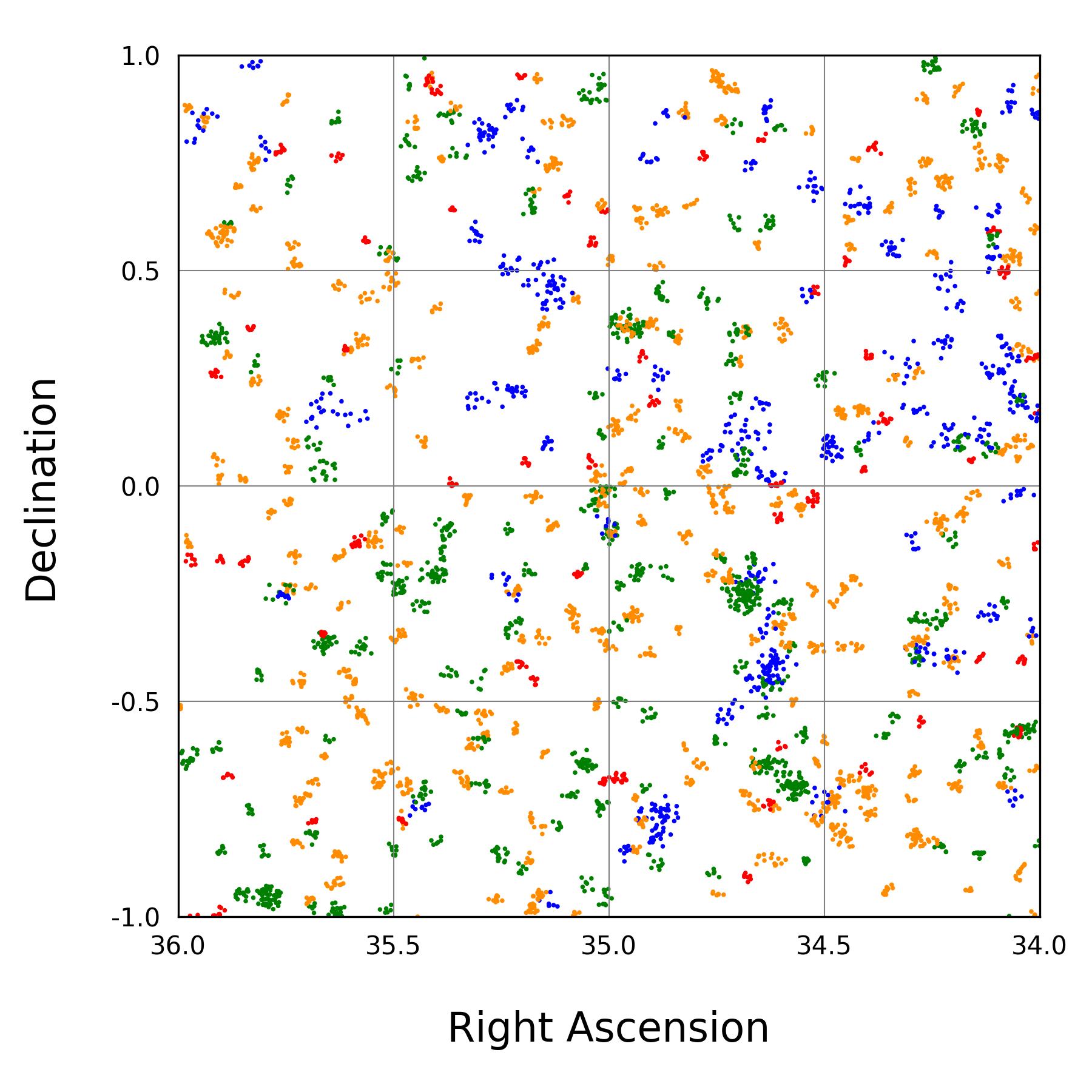}
\caption{Cluster member galaxies in a 2\textdegree $\times$ 2\textdegree portion of the WH24 catalog. Each point represents an individual galaxy. The different colors correspond to different redshift bins: blue is for $z < 0.4$, green is for $z = 0.4-0.7$, orange indicates $z = 0.7-1$, and red is for $z > 1$.}
\label{fig:fig3}
\end{figure}

The moment of inertia tensor is symmetrical, hence $I_{ij} = I_{ji}$. Diagonalizing this tensor yields the principal axes,
with the eigenvectors of the major axis giving the cluster's orientation and the ratio of the eigenvalues giving its ellipticity, $e$,

\[ e = 1 - \frac{\lambda_{\text{min}}}{\lambda_{\text{max}}} \]

\smallskip
\noindent where \( \lambda_{\text{max}} \) and \( \lambda_{\text{min}} \) are the maximum and minimum eigenvalues, respectively. Defined this way, \( e \) $\rightarrow 0$ for the roundest clusters, and $e \rightarrow$ 1 for the most elongated.
While the moment of inertia tensor allows for assigning a weight to each galaxy, such as its mass or luminosity, for simplicity we have chosen to weight each galaxy equally, treating each as identical 
tracers of the cluster's gravitational potential.

A second {\it independent} method was also used to measure cluster orientations and shapes based on the Random Sample Consensus (RANSAC) method \citep{RANSAC1981}. RANSAC is an iterative algorithm for estimating the parameters of a model that is robust against outliers. It works by repeatedly selecting random subsets of the data, fitting a model to them, and checking how well the model fits the rest of the data. This method is often used for detection of linear features in computer vision, geospatial analysis, and other fields. Comparison of the two methods showed excellent agreement, with a median absolute difference of only $15.3^{\circ}$ between cluster position angles determined from moments of inertia versus RANSAC, demonstrating the robustness of the orientation measurements. For simplicity, we present only the results based on moments of inertia in this paper.

We detected a very weak trend for the mean cluster ellipticities to increase with redshift. However, this is most likely explained 
 by the fact that detected cluster members are on average more massive and their number, $N_{gal}$, is smaller at higher redshifts
 (Fig. 2), which results in more elongated distributions statistically, a result we confirmed with simple simulations. 
When we calculated the mean ellipticity versus redshift 
using only clusters with $N_{gal} \leq 10$ members, hence comparing similar clusters at all redshifts (eliminating any 
dependence of ellipticity on $N_{gal}$), then the ellipticities are nearly constant.

In the absence of systematic errors, cluster orientations should be distributed uniformly across the sky.
We tested the null hypothesis that the distribution of observed cluster position angles is consistent with circular uniformity using a Bayesian model comparison. Both BIC-based and Laplace-based Bayes factors strongly favored the uniform model over a von Mises alternative of global asymmetries in cluster orientations. There is no evidence of systematic dependence of cluster orientations on Right Ascension or Declination (e.g., along equatorial strips).

Orientation uncertainties were estimated using bootstrap resampling with replacement.
The $1\sigma$ uncertainties range from $\sigma \sim 16$\textdegree\, for clusters with $N \geq 100$ member galaxies to $\sigma \sim 25$\textdegree\, for $N \le 10$.
Although individual cluster orientations may have significant uncertainties, the enormous number of clusters makes it feasible to tease out an alignment signal if present.
Moreover, the excellent agreement between cluster position angles determined using moments of inertia and RANSAC supports the overall reliability of these measurements.

\subsection{Cluster alignments}

The tendency for neighboring clusters to share similar orientations was measured by comparing the acute angle $\theta$ between each cluster’s major axis and the projected vector between its centroid and that of another cluster some distance, $D$, away. A total of 5.54 billion unique cluster pairs were found with comoving separations $d < 500$ cMpc.

If cluster orientations are uncorrelated, then $cos(2\theta)$ is expected to have a uniform random distribution between -1 and 1, with a mean value
$\langle cos(2\theta)\rangle = 0$. Positive values of $\langle cos(2\theta) \rangle$ indicate alignments, while  negative values correspond to anti-alignments. 

Figure 4 shows the alignments of WH24 clusters over different scales and redshift intervals. Statistical significance of alignments was assessed by generating 1000 Monte Carlo simulations in which each cluster was assigned a position angle from a uniform random distribution, and the entire analysis repeated. This accounts for any systematic effects that might result from the non-uniform sky coverage of the cluster sample. These results are also shown in Fig. 4. Similar results were obtained when the observed cluster position angles were shuffled rather than randomly assigned.

Alignments were considered significant if the observed $\langle \cos(2\theta) \rangle$ value exceeds the range of Monte Carlo values, indicating a probability $p < 0.001$ that the observed value is consistent with uncorrelated cluster orientations. Statistically significant correlations between cluster orientations are seen for separations as large as $\sim$ 200 - 250 cMpc for redshifts $z < 1$, corresponding to look-back times of 8 billion years.
For redshifts $z > 1$, statistically significant alignments are detected over pair separations up to $\sim 100$ cMpc.

An exponential function of the form $\langle \cos(2\theta) \rangle(d) = \alpha e^{-d/\tau}$
was found to provide a satisfactory empirical fit to the data, where $\alpha$ is the amplitude and $\tau$ is the characteristic decay scale. These fits are shown in Fig. 4, and the parameters of the best-fit exponential functions are given in Table 1.

\begin{figure*}[ht!]
\centering
\epsscale{0.9}
\includegraphics[width=1.0\textwidth]{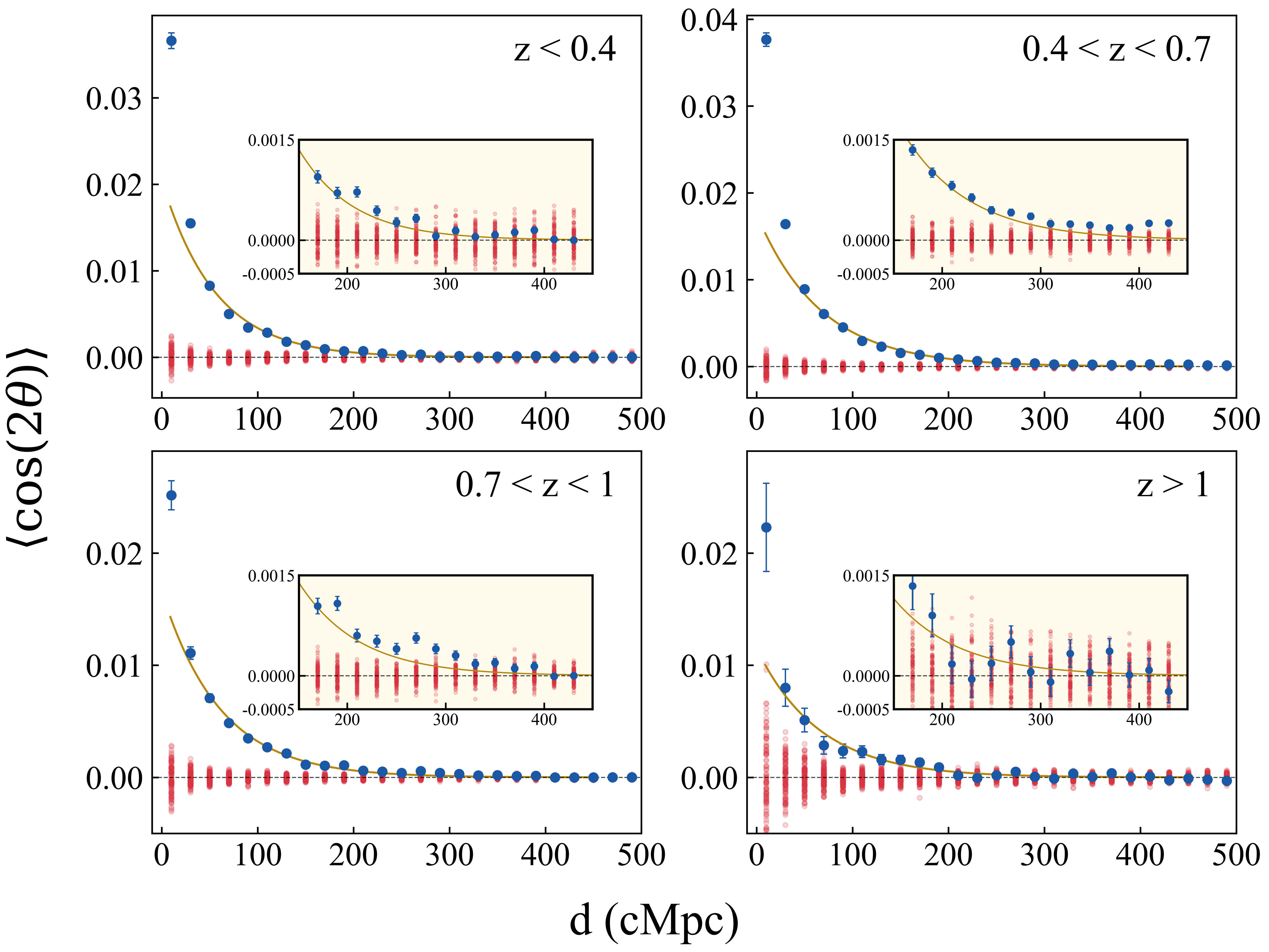}
\caption{Alignments of galaxy clusters in the WH24
sample as a function of comoving 
separation ($d$) and their evolution with the redshift ($z$). 
The statistic $\langle \cos(2\theta)\rangle$ measures the strength of cluster alignments, with an expectation value of $0$ if cluster orientations are uncorrelated. 
Error bars indicate the standard error of the mean for each bin.
The orange curve is a best-fit exponential function for each redshift bin; the amplitude ($\alpha$) and decay scale ($\tau$) of the best-fit components are listed in Table 1. 
The red dots are the results from each of 1,000 Monte Carlo simulations in which each cluster was assigned a random orientation, destroying any intrinsic alignments that might be present in the data.
The insets show an expanded view for the largest separations.
Statistically significant alignments are found over separations up to 
200-300 cMpc in all but the highest redshift subsample.
}
\label{fig:fig4}
\end{figure*}

\begin{table}[b!]
\centering
\caption{Best-fit exponential function parameters: {$\langle cos(2\theta)\rangle(d) = \alpha \exp\left({-d/\tau}\right)$ \\}}
\begin{tabular}{lll}
\toprule
Redshift & $\alpha$ & $\tau (cMpc)$ \\
$z < 0.4$ & $0.021 \pm 0.005$ & $55.5 \pm 7.4$ \\
$0.4 < z < 0.7$ & $0.018 \pm 0.007$ & $67.1 \pm 14.0$ \\
$0.7 < z < 1$ & $0.017 \pm 0.004$ & $60.9 \pm 7.8$ \\
$z > 1$ & $0.011 \pm 0.003$ & $65.6 \pm 11.0$ \\
\bottomrule
\end{tabular}
\end{table}

\subsection{Alignment dependence on cluster mass and ellipticity}

WH24 derived cluster masses based on the strong empirical correlation between cluster richness and mass within 
$r_{500}$, the projected radius within which the mean density of the cluster is 500 times the universe's critical density \citep{Wen2024, Wen2015}. 
Figure 5 shows that the most massive clusters tend to be most strongly aligned with their neighbors, perhaps indicating that they trace the richest superclusters. 

Recently, 
\citet{Einasto2024} showed that in the local Universe, high-mass clusters are located in filaments in high-density regions, in superclusters or in supercluster outskirts.
In such environments clusters grow by infall and merging of groups and galaxies 
along filaments, and alignments of clusters is a signature of this process. Poor groups can also be found in weak filaments, where  
the alignment signal is lower, as indicated by our results.

A weaker correlation between cluster ellipticity and alignment is also seen in Fig. 5, with the most elongated clusters showing the strongest alignment with their neighbors. This could reflect a tendency for the most elongated clusters to be most strongly aligned with anisotropic distribution of matter that surrounds them, or the increased uncertainties in measuring orientations of rounder clusters, or perhaps both.

\begin{figure*}[t!]
\centering
\includegraphics[width=0.9\textwidth]{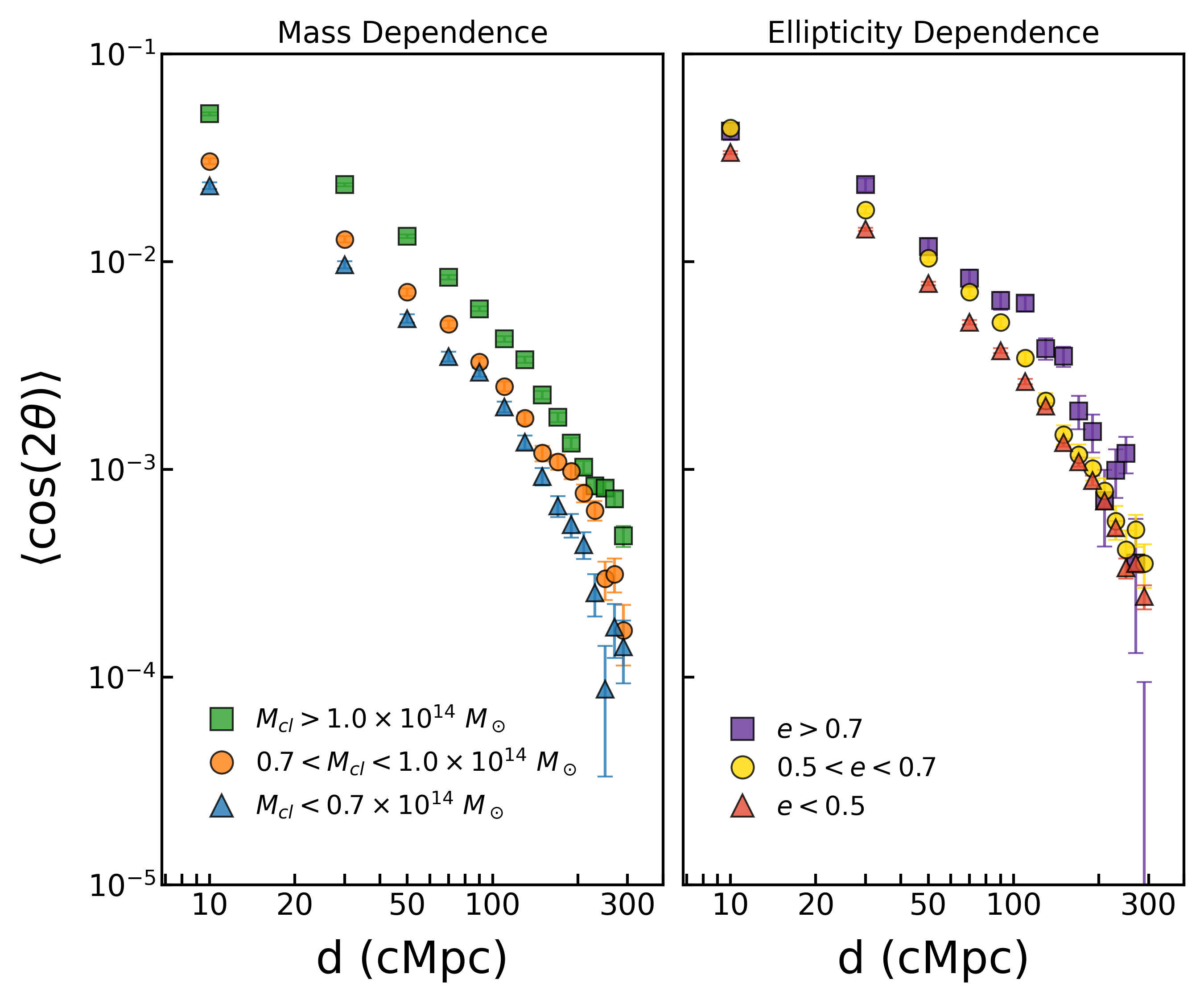}
\caption{Dependence of alignments on cluster mass (left) and ellipticity (right), for all redshifts combined. Note that in some cases the error bars are smaller than the symbol size.
The most massive and 
most elongated clusters are the most strongly aligned with their neighbors.
The data are plotted logarithmically to facilitate comparison.
}
\label{fig:fig5}
\end{figure*}

\subsection{Ruling out systematics}

To further ensure the reality of the alignments signal seen in Fig. 4, we examined subsets of the cluster data.

Although photometric redshifts of galaxies are inherently less precise than those determined from spectroscopy, the agreement is generally quite good. 
A comparison of photometric and spectroscopic redshifts from the Sloan Digital Sky Survey (SDSS) Data Release 6 
showed that the completeness of member galaxies was about 90\%, and the contamination rate was about 20\% \citep{Wen2009}. WH24 did the same for DESI galaxies using spectroscopic redshifts from the deeper 
Galaxy And Mass Assembly Survey \citep{Liske2015,Driver2022} and found similar results. 
Thus, most member galaxies found using photometric redshifts are genuine. 
Furthermore, because cluster redshifts are derived from multiple member galaxy candidates, their uncertainties are smaller
than those of individual galaxies. As shown in Fig. 7 of WH24, the uncertainty of ($z_{spec} - z_{phot})/(1 + z_{spec}$) is less than 0.01 at $z < 0.6$ 
and increases to 0.019 at $z \sim 1.15$ for the clusters with i-band data (in the region of declination $<32$\textdegree). For the clusters
without i-band data (most in the region at declination $> 32$\textdegree), the uncertainty of cluster redshift is slightly larger,
about 0.01 at $z<0.6$, and increases to 0.025 at $z \sim 1.15$.
This translates into uncertainties of a few tens of comoving Megaparsecs (cMpc) in cluster distances. Although such uncertainties introduce noise into our analysis, it only diminishes the ability to detect intrinsic cluster alignments and cannot create a false positive detection. Moreover, the enormous number of clusters in the sample helps to ameliorate the impact of individual distance uncertainties.
We note that the WH2024 catalog adopts spectroscopic redshifts when available, with the spectroscopic redshift of the 
brightest member galaxy usually taken as the cluster redshift.

To demonstrate that our use of photometric redshifts does not produce a spurious signal of cluster alignments, we repeated the analysis using only the 21\% of WH24 clusters whose redshifts have have been measured spectroscopically. Despite the even smaller sample size, statistically significant alignments are again seen for $z < 1$, while at higher redshifts, the results are consistent with no alignments. 
As discussed earlier, this might indicate weaker alignments at those redshifts or it might simply be reflect small number statistics resulting from the very small number of clusters with spectroscopic redshifts at $z > 1$.
Of the 139,527 clusters with redshifts $z > 1$ in the WH24 catalog, fewer than 1\% have been obtained spectroscopically. 
Consequently, the much small number of cluster pairs in this subsample results in quite large uncertainties.

Because the reliability of a cluster’s measured orientation typically increases with the number of member galaxies, we repeated the analysis using only clusters with 10 or more detected members.
This subsample comprises 35\% of the entire WH24 catalog. The statistical trade-off, of course, is far fewer clusters but with more 
robust orientation measurements. 
Despite the smaller sample size, statistically significant alignments are still detected for pair separations in excess of $\sim 100$ cMpc.

\section{Comparison with cosmological simulations} 

Additional insights can be gained by examining cluster alignments predicted by numerical simulations. For this purpose, we used the high-resolution, gravity-only LastJourney simulation \citep{Habib2016,Heitmann2019,Heitmann2021}, which follows the evolution of more than a trillion particles within a $(5.025$ cGpc$)^3$ cubic periodic volume starting from initial conditions consistent with the Planck $\Lambda$CDM cosmology. The mass of each particle is $4 \times 10^{9}\solarmass$. Data are publicly available for redshifts between $z=0$ and $z=1.5$, and gravitationally bound halos have been identified using a friends-of-friends algorithm (see the aforementioned papers for details).

We selected simulated clusters with halo masses M$_{cl} \geq 5 \times 10^{14}\solarmass$ at redshift $z = 0$, corresponding to typical rich clusters in the present-day universe and in the WH24 catalog. A total of 105k such clusters were found at this redshift. Because the average cluster mass is smaller at earlier epochs, we chose a slightly lower mass cutoff of M$_{cl} \geq 2.5 \times 10^{14}\solarmass$ for $z = 0.5$, and M$_{cl} \geq 1.5 \times 10^{14}\solarmass$ for $z = 0.9$ and 1.5.

Once cluster halos were selected, their orientations and ellipticities were calculated from the distribution of their member particles using the same moments-of-inertia method described earlier. Although the LastJourney simulation is fully three-dimensional, cluster orientations and alignments were measured using three orthogonal, two-dimensional projections in order to compare more directly with the observations,
resulting in a sample that is effectively three times larger. A few examples of simulated clusters are shown in Fig. 6. 
Each cluster has thousands of particles, yielding much more robust major axis position angles compared to the observed clusters. 
Bootstrap resampling shows that the simulated clusters' orientations have uncertainties less than one degree. 

The dot product of the two-dimensional vector defined by a cluster's major axis and the projected vector between it and a neighboring cluster a distance $D$ away was computed for all simulated cluster pairs. The minimum image convention was used to calculate distances given the periodic boundaries of the simulated volume.

Figure 7, where we plot the alignments in the simulations over different scales and redshifts, shows clearly that cluster alignments are present in the simulations.
As with the observed clusters, we consider alignments to be significant if $\langle \cos(2\theta) \rangle$ has a probability $p < 0.001$ of occurring according to the Monte Carlo simulations with random cluster orientations. The simulated clusters show statistically significant alignments over separations up to $\sim 200-300$ cMpc for redshifts $z < 1$, comparable to the observed clusters, and to $\sim 50$ cMpc at $z = 1.5$. The amplitude of these alignments is stronger than observed in Fig. 4, however, as noted by \cite{Smargon2012}, this likely reflects the idealized nature of simulations and the larger uncertainties in the observational data.

\begin{figure*}[ht!]
\centering
\includegraphics[width=1.0\textwidth]{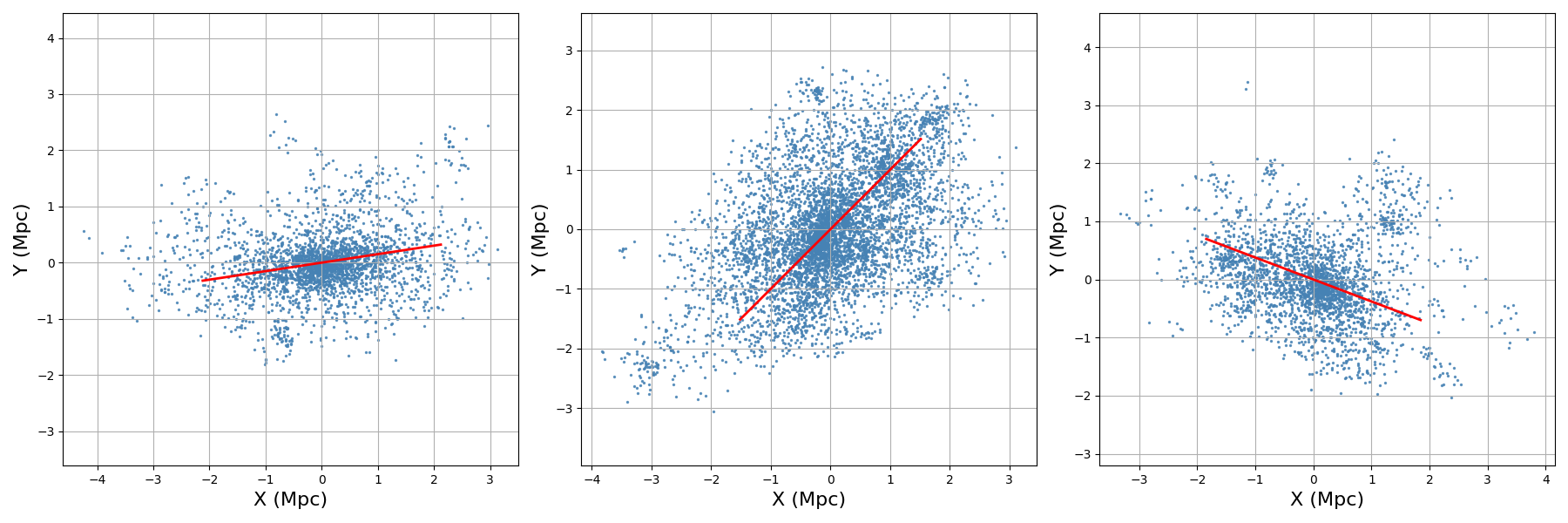}
\caption{Projected particle distribution in three typical LastJourney clusters at redshift $z = 0$. The major axis orientation for each, determined from the moments of inertia of its particle distribution, is shown in red. Coordinates are relative to the cluster center.}
\label{fig:fig6}
\end{figure*}

\begin{figure*}[ht!]
\centering
\includegraphics[width=0.9\textwidth]{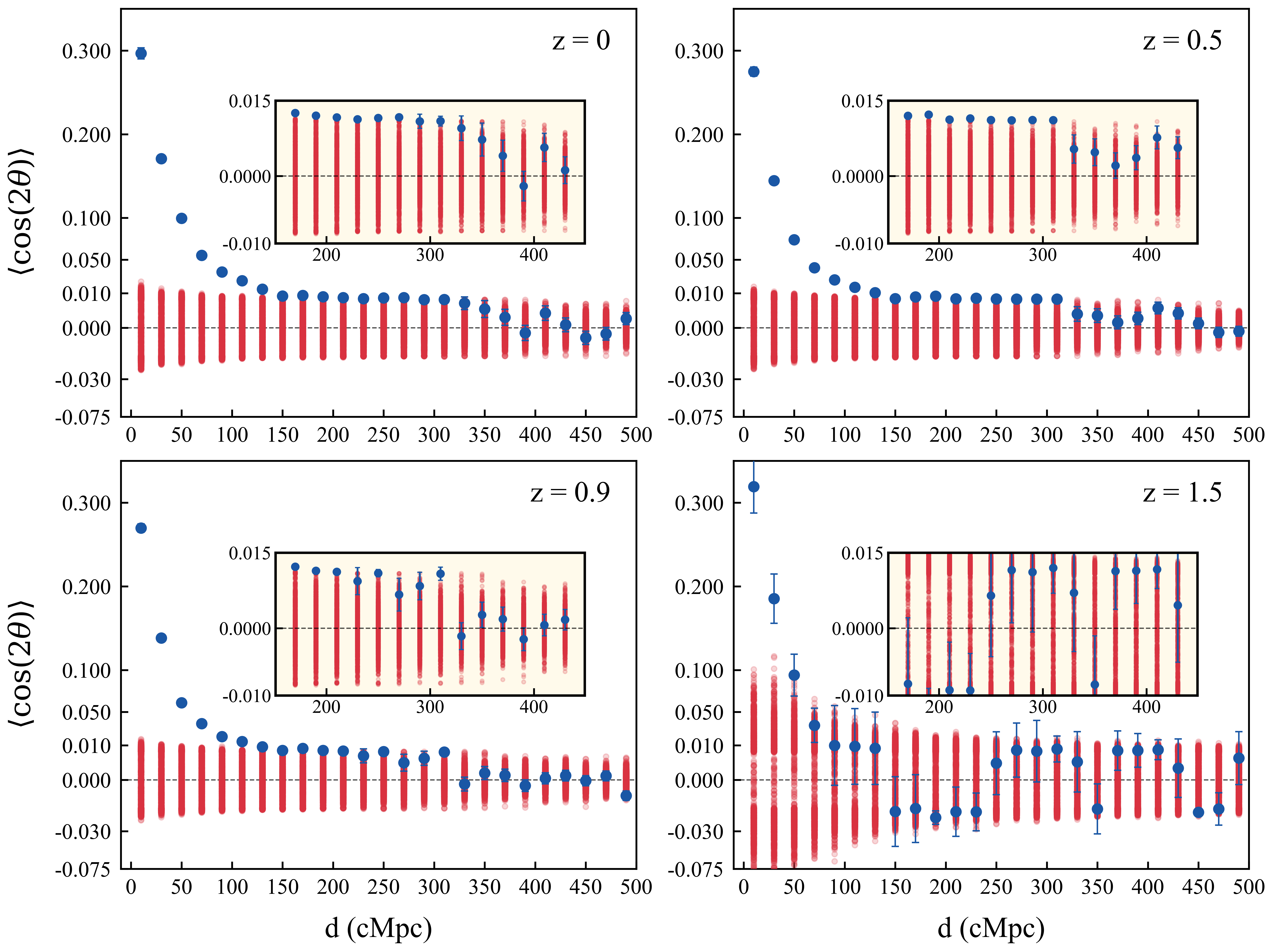}
\caption{Cluster alignments at four different redshifts in the LastJourney simulation.
As in Fig. 4, the red dots are the results from 1,000 Monte Carlo simulations in which each cluster was assigned a random orientation.
The insets show an expanded view for the largest separations.
Statistically significant alignments are found in the simulations for separations up to 
200-300 cMpc in all but the highest redshift clusters.}
\label{fig:fig7}
\end{figure*}

\section*{Discussion}

We have shown for the first time that correlations between cluster orientations are observed over scales of $200-300$ cMpc or more and out to redshifts $z > 1$, spanning a majority of the universe's history. This indicates that
the formation of these Mpc-scale systems has been influenced by the anisotropic distribution of matter on scales two orders of magnitude greater than their characteristic sizes, most likely via anisotropic accretion of material
along filaments 
(see also \citeauthor{West1995} \citeyear{West1995};
\citeauthor{Wen2024b} \citeyear{Wen2024b};
\citeauthor{Einasto2024} \citeyear{Einasto2024}).
Our results demonstrate that cluster alignments are a powerful probe of the cosmic web at earlier epochs. 

These results suggest that the ``skeleton'' of the cosmic web (the seeds of superclusters) formed in the very early universe and that the essential evolution of clusters occurs within supercluster environments where clusters grow by the infall of galaxies, groups and neighboring clusters along filaments \citep{Einasto2019,Chiang2013,Einasto2021,Einasto2024}. 
It also suggests that alignments are likely to be observable at high redshifts
as larger samples of distant clusters become available.  
However, the challenge of identifying clusters and protoclusters at early epochs remains formidable.

The coherence scale of the observed cluster alignments is comparable to that found in the LastJourney simulations of a canonical $\Lambda$CDM cosmology. Enormous, flattened superstructures of galaxies (superclusters, supercluster complexes, and planes) spanning many hundreds of Mpc are known to exist in the local universe. The largest among such systems are the Local Supercluster plane and perpendicular to it the Dominant Supercluster Plane, delineated by galaxies, clusters and rich superclusters \citep{Einasto1983,Tully1986,Einasto1994,Einasto1997,Peebles2022,Peebles2023}. In addition, rich superclusters form complexes with length over two hundred megaparsecs in which several rich superclusters are close and aligned, as in the Coma Wall, in the Sloan Great Wall, and in the
BOSS Great Wall at redshift $z \approx 0.46$   \citep{deLapparent1986,Gott2005,Lietzen2016,Einasto2016, Einasto2022}. 

The existence of such large structures and correlated alignments of galaxy clusters 
raises intriguing questions about the scale of homogeneity in the universe and whether 
they violate the standard $\Lambda$CDM model and the cosmological principle. 
\cite{Peebles2023} recently argued that huge supercluster planes could pose problems 
for $\Lambda$CDM. 
Some authors suggest that inhomogeneities are not expected to exist on scales larger 
than a few hundred Mpc \citep{Yadav2010,Planck2016,Melia2023}, although see 
\cite{Park2012} and \cite{Kim2022}. 
However, the sizes of the largest structures in the local Universe exceed this value 
\citep{Einasto1997,Lietzen2016}.
\cite{Boehringer2025} recently surveyed large-scale structures in the local universe using X-ray clusters and identified a superstructure on a scale of 400 Mpc that they named Quipu. 
This very elongated superstructure consists of three poor superclusters \citep{Einasto2001}, connected by a filament of X-ray groups and clusters. 
The Quipu may represent a structure with correlated alignments of clusters along its axis, but this, of course, needs a separate study.

New studies based on CosmicFlows4 data demonstrate that within distances of  
$200-300$ Mpc from the observer homogeneity is not yet reached \citep{Courtois2025}.
This result is in agreement with the presence of superclusters and supervoids
with sizes over $\sim 300$ Mpc \citep{Einasto2011,Liivamagi2012,Park2012}.

Although even larger structures have been reported \citep{Clowes2013,Horvath2014,Balazs2015,Lopez2022,Lopez2024}, the reality of these features remains controversial (e.g., \cite{Scrimgeour2012,Einasto2014,Park2015,Kumar2023,Fujii2024,Goyal2024}). \cite{Sawala2025} further suggest that such apparent patterns 
are common random configurations and likely reflect the algorithms used to identify them rather than 
genuine mass concentrations. 

The scale of homogeneity 
must of course decrease linearly with redshift in the matter-dominated era as structures grow. \cite{Goncalves2018,Goncalves2021} use SDSS quasars to estimate
that the bias corrected homogeneity scale at $z=1$ is about 120 Mpc for our adopted value of the Hubble constant, 75 Mpc when corrected for bias) which is claimed to be in good agreement with $\Lambda$CDM predictions. This is somewhat smaller than the scale 
at which the alignment signal disappears in Fig. 4, but at the 1\% 
inhomogeneity level superclusters might exceed this measurement. Observations at higher redshifts would provide a more stringent
limit, as the homogeneity scale decreases linearly.

Finally, we note that cluster alignments complement other diagnostics of the cosmic web's evolution, such as weak lensing \citep{Dietrich2012,HyeongHan2024}, quasars \citep{Einasto2014,Wang2023}, Lyman-$\alpha$ emitters \citep{Umehata2019,Martin2023} and cosmological 21 cm emission \citep{Amiri2023}. 
To understand the evolution of galaxy clusters in the cosmic web, and the origin and evolution of cluster alignments in detail, data from various current and future surveys such as J-PAS \citep{Benitez2014},
4MOST \citep{DeJong2019}, and, of course, new DESI data releases \citep{Dey2019} can be employed.

\section{Conclusions}

Using the largest catalog of galaxy clusters currently available \citep{Wen2024}, with 1.58 million clusters, we have examined the evolution of cluster alignments over billions of years and compared the results with numerical simulations of a $\Lambda$CDM universe \citep{Heitmann2021}. 
Cluster orientations are found to be correlated over scales up to $\sim 200-300$ cMpc in both observed and simulated clusters. This is true for the present day and out to redshifts of at least $z\simeq 1$, corresponding to look-back times of more than half the age of the universe.
These results suggest that cluster and proto-cluster alignments may provide a useful probe of the cosmic web's development at even higher redshifts.

\section{Acknowledgements}
We thank the anonymous referee and the statistics editor for helpful feedback that improved this paper, and also Hans B\"ohringer and Gayoung Chon for helpful discussions. MW thanks the Finnish Centre for Astronomy with ESO (FINCA), the University of Turku, the U.S. Fulbright program, and the Fulbright Finland Foundation for their support and hospitality during this research. 
ME acknowledges support by 
ETAG CoE grant “Foundations of the Universe" (TK202), by ETAG grant PRG2172,
and the European Union's Horizon Europe research and innovation programme 
(EXCOSM, grant No. 101159513).
ZLW and JLH are supported by the National Natural Science Foundation
of China (Grant Numbers 11988101, 11833009 and 12073036), the Key
Research Program of the Chinese Academy of Sciences (Grant Number
QYZDJ-SSW-SLH021) and also the science research grants from the China Manned Space Project with Numbers CMS-CSST-2021-A01 and CMS-CSST-2021-B01.

The Legacy Surveys consist of three individual and complementary projects: the Dark Energy Camera Legacy Survey (DECaLS; Proposal ID \#2014B-0404; PIs: David Schlegel and Arjun Dey), the Beijing-Arizona Sky Survey (BASS; NOAO Prop. ID \#2015A-0801; PIs: Zhou Xu and Xiaohui Fan), and the Mayall z-band Legacy Survey (MzLS; Prop. ID \#2016A-0453; PI: Arjun Dey). DECaLS, BASS and MzLS together include data obtained, respectively, at the Blanco telescope, Cerro Tololo Inter-American Observatory, NSF’s NOIRLab; the Bok telescope, Steward Observatory, University of Arizona; and the Mayall telescope, Kitt Peak National Observatory, NOIRLab. Pipeline processing and analyses of the data were supported by NOIRLab and the Lawrence Berkeley National Laboratory (LBNL). The Legacy Surveys project is honored to be permitted to conduct astronomical research on Iolkam Du’ag (Kitt Peak), a mountain with particular significance to the Tohono O’odham Nation.

NOIRLab is operated by the Association of Universities for Research in Astronomy (AURA) under a cooperative agreement with the National Science Foundation. LBNL is managed by the Regents of the University of California under contract to the U.S. Department of Energy.

This project used data obtained with the Dark Energy Camera (DECam), which was constructed by the Dark Energy Survey (DES) collaboration. Funding for the DES Projects has been provided by the U.S. Department of Energy, the U.S. National Science Foundation, the Ministry of Science and Education of Spain, the Science and Technology Facilities Council of the United Kingdom, the Higher Education Funding Council for England, the National Center for Supercomputing Applications at the University of Illinois at Urbana-Champaign, the Kavli Institute of Cosmological Physics at the University of Chicago, Center for Cosmology and Astro-Particle Physics at the Ohio State University, the Mitchell Institute for Fundamental Physics and Astronomy at Texas A\&M University, Financiadora de Estudos e Projetos, Fundacao Carlos Chagas Filho de Amparo, Financiadora de Estudos e Projetos, Fundacao Carlos Chagas Filho de Amparo a Pesquisa do Estado do Rio de Janeiro, Conselho Nacional de Desenvolvimento Cientifico e Tecnologico and the Ministerio da Ciencia, Tecnologia e Inovacao, the Deutsche Forschungsgemeinschaft and the Collaborating Institutions in the Dark Energy Survey. The Collaborating Institutions are Argonne National Laboratory, the University of California at Santa Cruz, the University of Cambridge, Centro de Investigaciones Energeticas, Medioambientales y Tecnologicas-Madrid, the University of Chicago, University College London, the DES-Brazil Consortium, the University of Edinburgh, the Eidgenossische Technische Hochschule (ETH) Zurich, Fermi National Accelerator Laboratory, the University of Illinois at Urbana-Champaign, the Institut de Ciencies de l’Espai (IEEC/CSIC), the Institut de Fisica d’Altes Energies, Lawrence Berkeley National Laboratory, the Ludwig Maximilians Universitat Munchen and the associated Excellence Cluster Universe, the University of Michigan, NSF’s NOIRLab, the University of Nottingham, the Ohio State University, the University of Pennsylvania, the University of Portsmouth, SLAC National Accelerator Laboratory, Stanford University, the University of Sussex, and Texas A\&M University.

BASS is a key project of the Telescope Access Program (TAP), which has been funded by the National Astronomical Observatories of China, the Chinese Academy of Sciences (the Strategic Priority Research Program “The Emergence of Cosmological Structures” Grant \# XDB09000000), and the Special Fund for Astronomy from the Ministry of Finance. The BASS is also supported by the External Cooperation Program of Chinese Academy of Sciences (Grant \# 114A11KYSB20160057), and Chinese National Natural Science Foundation (Grant \# 12120101003, \# 11433005).

The Legacy Survey team makes use of data products from the Near-Earth Object Wide-field Infrared Survey Explorer (NEOWISE), which is a project of the Jet Propulsion Laboratory/California Institute of Technology. NEOWISE is funded by the National Aeronautics and Space Administration.

The Legacy Surveys imaging of the DESI footprint is supported by the Director, Office of Science, Office of High Energy Physics of the U.S. Department of Energy under Contract No. DE-AC02-05CH1123, by the National Energy Research Scientific Computing Center, a DOE Office of Science User Facility under the same contract; and by the U.S. National Science Foundation, Division of Astronomical Sciences under Contract No. AST-0950945 to NOAO.

For the LastJourney simulations, an award of computer time was provided by the ASCR Leadership Computing Challenge (ALCC) program. This research used resources of the Argonne Leadership Computing Facility at Argonne National Laboratory, which is supported by the Office of Science of the U.S. Department of Energy under contract No. DE-AC02-06CH11357.

\bibliography{West}
\bibliographystyle{aasjournal}



\end{document}